\documentstyle[prd,aps,epsf,floats,twocolumn]{revtex}
\begin{document}
\draft

\title{FRW cosmologies between chaos and integrability}
\author{A. E. Motter$^{a}$ and P. S. Letelier$^{b}$}
\address{Departamento de Matem\'atica Aplicada - IMECC,  
Universidade Estadual de Campinas (UNICAMP), 13083-970 Campinas, Brazil}
\date{Phys. Rev. D {\bf 65}, 068502 (2002)}
\maketitle

\begin{abstract}
A recent paper by Castagnino, Giacomini and Lara concludes
that there is no chaos in a conformally coupled closed
Friedmann-Robertson-Walker universe, which is in apparent contradiction with
previous works.
We point out that although nonchaotic the quoted system is nonintegrable.
\end{abstract}


Castagnino, Giacomini and Lara (CGL) \cite{castagnino} analyzed the
dynamical evolution of the spatially closed Friedmann-Robertson-Walker (FRW)
cosmology conformally coupled to a massive, real, scalar field. Employing the
cosmological time as the time parameter, they show that for arbitrary initial
conditions the universe will collapse in a finite time and with a divergent
rate of contraction. Their conclusion is that there is no chaos in the model.

The same system had been previously considered by Calzetta and El Hasi
\cite{calzetta1}, who presented evidence of chaotic behavior
subsequently confirmed by Bombelli, Lombardo and Castagnino \cite{bombelli}.
Since these authors employed the conformal time in their analyses,
CGL argue that the discrepancy
of the results would rely on the different time parameters used -- the
origin of an intensive debate in the paradigmatic mixmaster model
\cite{motter1}. The aim of this note is to show that this is not the case
and that all these fine works are in complete agreement in spite of the
different choices of coordinates.

The classical theory of dynamical systems regards the study of systems of the
form
\begin{equation}
\frac{d{\mathbf x}}{dt}=\bf{F}({\mathbf x}),
\label{classical}
\end{equation}
for a fixed choice of the time parameter $t$. An invariant set of the phase
space ${\mathbf x}$ is chaotic if it presents a {\it sensitive dependence on
initial
conditions and mixing}. This characterization is invariant under space
diffeomorphisms: ${\mathbf y}={\mathbf \psi}({\mathbf x})$. In general
relativity, the absence of an absolute time forces us to consider system
(\ref{classical}) under space-{\it time} diffeomorphisms:
${\mathbf y}={\mathbf \psi}({\mathbf x},t)$, $d\tau=\lambda({\mathbf x},t)dt$.
Usually, the dynamical variables in this context are either functions of the
space-time coordinates $x^{\mu}$
(possibly together with spins, Euler angles, etc.), when we study motions in a
given background geometry, or functions of the metric $g_{\mu\nu}$
(possibly together
with other fields), when we consider the evolution of the geometry itself. The
study of chaos in general relativity faces both conceptual and technical
difficulties. The former are associated with the dependence of classical
indicators of chaos on the choice of the time parameter. This problem has been
intensively discussed in the literature since Francisco and Matsas
\cite{francisco} showed the coordinate dependence of Lyapunov exponents.
The latter
difficulties are related with characteristic properties of relativistic
systems. In cosmology, for example, we often meet high dimensionality,
noncompacity, nonpositive kinetic energy, non-normalizable measure,
nonexistence of global coordinates, nontrivial topology, singularities, etc.
These properties strongly restrict the practical use of standard indicators of
chaos, even if the system is treated as a classical one. The second class of
difficulties is in fact the origin of most of the problems concerning chaos in
cosmology, including the one discussed in this communication.

In terms of the conformal time $\eta$ the system considered by
CGL \cite{castagnino} is modeled by the $H=0$ energy surface of the
Hamiltonian \begin{equation}
2H=-p_a^2+p_{\phi}^2-a^2+{\phi}^2+a^2{\phi}^2,
\label{hamilt}
\end{equation} 
where $a$ is the radius of the universe and $\phi$ is the reparametrized scalar
field \cite{footnote}. CGL proved that the dynamics is nonchaotic {\it if}
formulated in terms of the cosmological time. Our first point is that this is
also true when the dynamics is formulated in terms of the conformal time as
well as in terms of any other well defined time parameter. Translating the CGL
work into conformal time it follows that for every physical initial condition,
$a>0$ at $\eta=0$, there will be a finite time $\mathnormal{\eta_1}$ such that
 $a\rightarrow 0$ and 
$a'\rightarrow C$ for $\eta \rightarrow\mathnormal{\eta_1}$,
where the prime denotes $d/d\eta$ and
$C$ is a nonzero, negative, finite constant. The absence of physical
meaning for a negative universe radius, $a<0$, prevents us from extending the
solutions beyond the big crunch. Since chaos is a concept associated with an
infinite number of recurrences, it is clear that (\ref{hamilt}) regarded as a
cosmological system is nonchaotic. To see the generality of this statement we
can consider the Ricci scalar
\begin{equation}
\frac{R}{6}=m^2\left(\frac{\phi}{a}\right)^2,
\label{ricci} 
\end{equation}
where $m$ is the mass of the coupled field. 
From Eq. (\ref{hamilt}) we have $\phi '^2+\phi^2\rightarrow C^2$ when
$a\rightarrow 0$ and  $a'\rightarrow C$, which leads to two possibilities:
$R\rightarrow 6m^2$ if $\phi\rightarrow 0$, and  $R\rightarrow \infty$ if
$\phi\rightarrow {\mathnormal \phi_1}\neq 0$. It is easy to see from a
Poincar\'e section defined by $a=0$, $a'<0$ that $\phi$ is typically nonzero
at the big crunch, resulting in a divergent  behavior for $R$. Such singularity
is coordinate invariant and forbids physical extensions of the solutions
throughout the big crunch, whatever coordinates we use.   

Since there is no chaos the next question is whether or not the system is integrable.
The answer comes from Refs. \cite{calzetta1,bombelli}, where it is
shown that the universe would present chaotic regions in the phase space
$(p_a,p_{\phi},a,\phi)$ for an {\it infinite} sequence of contractions and
expansions \cite{melnikov}. This extension through $a<0$ has no physical
meaning, as mentioned before, but is an ingenious mathematical trick to obtain
properties in the physical region from properties in the extended, unphysical domain. For
example, chaos in the extended domain implies nonintegrability. In
particular, it implies nonintegrability in the physical region defined by the
first half-cycle $a>0$, between a big bang and the following big crunch. We
stress that, even though the extension of the dynamics beyond the big crunch can
be performed in the conformal time formulation and not in the cosmological time
approach (because of the  divergence of the contraction rate), the result
concerning nonintegrability in the physical region is invariant under
coordinate changes. In fact, the relation between the cosmological
time and the conformal time in the physical domain is $dT=a d\eta$, which is a
particular case of an {\it autonomous} transformation of the form 
\begin{eqnarray}
{\mathbf y}&=&{\mathbf \psi}({\mathbf x}),
\label{space}\\ 
d\tau &=&\lambda({\mathbf x})dt,
\label{time}
\end{eqnarray} 
where $\lambda$ is a positive function and ${\mathbf \psi}$ is a diffeomorphism.
The integrability is coordinate
invariant under this class of transformations because if $\{I_1,I_2,...\}$ 
are independent integrals of  motion in
the original variables $({\mathbf x},t)$  (i.e., ${\mathbf F}\cdot\nabla I_i=0$ 
and $\{\nabla I_1, \nabla I_2, ...\}$ are linearly independent) then
$\{I_1\circ{\mathbf \psi}^{-1}, I_2\circ{\mathbf \psi}^{-1},...\}$ are 
independent integrals of motion in the variables $({\mathbf y},\tau)$.
Therefore we can say that the conformally coupled FRW model (\ref{hamilt})
is nonintegrable in a meaningful sense.  Accordingly, we cannot hope to
find exact solutions for arbitrary initial conditions.

The same idea can be used to study the integrability of others FRW
cosmologies, as long as we can find nonsingular coordinates to mathematically
extend the solutions. In particular, this procedure works in the spatially
closed cosmology conformally coupled to a scalar field, with both mass and
cosmological constant terms, considered in Ref. \cite{calzetta2}. Since it was
shown that this model is chaotic in the extended domain, it follows that the
system is nonintegrable in the physical region. We observe that methods 
based on extensions to unphysical values have been used for a long time to
study integrability in cosmology. Perhaps the best known of them is the one
based on the  Painlev\'e theory of differential equations \cite{helmi}. In
the Painlev\'e analysis we look for necessary conditions for integrability
(equivalently, sufficient conditions for nonintegrability)
by studying critical points in the complex plane of time.

Summarizing, the dynamics of the model studied by CGL is chaotic when
analyzed for an unphysical sequence of expansions and contractions of the
universe, and is nonchaotic when considered for the period of time limited by
a big bang  and a big crunch. Nevertheless, it does not mean that the dynamics
is simple since the onset of chaos in the extended domain implies
nonintegrability in the physical region, which poses obstructions to the
study of exact solutions. This characterization is coordinate
invariant and does not rely on {\it natural} or {\it
physical} choices of the time parameter, consistent with the covariant
principle of general relativity.

A.E.M. thanks Y.-C. Lai for his kind hospitality at Arizona State University.
This work was supported by Fapesp and CNPq.


\begin{references}
\bibitem{footnote}
The dynamical variables have been rescaled by the mass 
of the field: $a\rightarrow a/m$ and $\phi\rightarrow\phi/m$.
\bibitem{melnikov}
The analysis of Hamiltonian (\ref{hamilt}) is highly non-trivial because
the kinetic energy $2K=-p_a^2+p_{\phi}^2$ is not positive definite.
We cannot find confined regions by analyzing the potential
$2V=-a^2+\phi^2+a^2\phi^2$, what makes difficult the use of standard methods
for chaos like Poincar\'e section and Lyapunov exponent methods. The authors
of Refs. \cite{calzetta1,bombelli} managed to apply Melnikov methods to study
the effect of the coupled field. For small values of $a$ and $\phi$, the
coupling term $a^2\phi^2/2$ is a small perturbation and only a small fraction
of the Kolmogorov-Arnold-Moser (KAM) tori are destroyed. For large values of
$a$ and/or  $\phi$, the coupling term becomes significant and most of the KAM
surfaces are destroyed, leading their orbits to access a higher dimensional
region.
\end{references}
\end{document}